\documentclass[review]{elsarticle}

\usepackage{lineno,hyperref}
\usepackage{xcolor}
\modulolinenumbers[5]
\usepackage{graphicx}
\usepackage{times}
\usepackage{textcomp}
\newcommand{\bef}{\begin{figure}}
\newcommand{\eef}{\end{figure}}
\newcommand{\bc}{\begin{center}}
\newcommand{\ec}{\end{center}}
\modulolinenumbers[5]

\journal{Astroparticle Physics}

\begin{document}
\begin{frontmatter}

\title{Cosmic ray measurements using the ISMRAN setup in a non-reactor environment}

\author[a,1]{S. P. Behera}
\ead{shiba@barc.gov.in}
\author[a,b]{R. Sehgal,}
\author[a,b]{R. Dey,}
\author[a]{P. K. Netrakanti,}
\author[a]{D. K. Mishra,}
\author[a,b]{V. Jha, and}
\author[a,b]{L.M. Pant}

\address[a]{Nuclear Physics Division, Bhabha Atomic Research Centre,\\ Mumbai - 400085, India}
\address[b]{Homi Bhabha National Institute, Anushakti Nagar, Mumbai - 400094, India}


\begin{abstract}
The cosmic rays data collected using a large area plastic scintillator 
array ISMRAN (Indian Scintillator Matrix for Reactor AntiNeutrino) are presented. 
The data collected serve as a useful benchmark of cosmogenic background in a 
non-reactor environment for the future measurements of electron-antineutrinos
 to be performed using the ISMRAN setup. The zenith angle distribution of the
 atmospheric muons has been measured and compared with Monte Carlo expectations.
The detector setup was further used to measure the lifetime distribution of 
stopped muons and extract their rates inside the detector matrix. The measured
 spectra of decaying muons and associated electrons show a good agreement with
 the MC simulations performed through GEANT4 simulation.
\end{abstract}

\begin{keyword}
\texttt{}Cosmic rays, scintillators, muons measurement

\end{keyword}

\end{frontmatter}

\section{Introduction}
Experimental measurements of cosmic-rays at sea level are important for 
studying the interaction of processes associated with its production. Further, 
data on cosmic-ray intensities serve as a useful benchmark for various other
astrophysical phenomena.
Cosmic muons are produced when primary high energetic cosmic particles interact
 in the upper atmosphere of the earth, producing secondary particles, such as 
charged pions, which in turn decay to produce muons and neutrinos via the 
processes $\pi^{-}\rightarrow \mu^{-} + \overline\nu_{\mu}, \pi^{+}\rightarrow \mu^{+} + \nu_{\mu}  $.
 Therefore, the measurements of cosmic muons provide a useful tool for the determination 
of calculating the atmospheric neutrino flux. The primary cosmic rays reaching on earth
 are more or less isotropic. On the other hand, cosmic muons flux at sea level depends
 on geomagnetic latitude, altitude, solar activity, and atmospheric condition. There are
 several parametrizations available for the angular and energy distribution of cosmic 
muons at sea level \cite{gassier}. The angular distribution of muons at various altitudes and latitudes
 follows the expression $I = I_{0}~cos^{n}~\theta$, where $I_0$ is the
 vertically integrated muon flux (cm$^{-2}$ s$^{-1}$ sr$^{-1}$) and
  $\theta$ is the zenith angle of incoming muons~\cite{pdg0,nima}.
  Both the quantities, the exponent $n$ and $I_0$ depend on momentum cut off, latitude, 
and altitude of incident muons. Several measurements have been carried out to find the 
angular and energy dependence of flux at sea level and charge ratio of cosmic muons 
which help to know the primary cosmic rays in more details~\cite{Aguilar:2010kg}.
 
The flux of muons at sea level is $\approx$ 1 per cm$^2$ per minute with mean energy
 of about 4 GeV. Due to high energy,  muons travel almost at the speed of light resulting
 in a transit time of about 106 $\mu s$ which is high compared to its decay time 
(mean lifetime $ \tau = 2.1969811 \pm 0.0000022~\mu s$)~\cite{pdg1} at rest. Most of
 the cosmic muons arriving on earth's surface are produced at around 15 km above 
sea level. As a result, most of the muons reach on earth before decaying in the atmosphere.
 The specific energy loss for muons is about 2 MeV per g/cm$^2$~\cite{pdg}, which is
 very small compared to the energy (GeV) of muons. Therefore, muons do not lose much
 energy in the medium as they travel, allowing them to penetrate more deeply into 
materials. Thus, the energy loss for a specific traversing distance can be determined,
 which is then used to calibrate the detector systems. Muons are not only used for
 studying the fundamental interactions but also used for societal applications such
 as muon radiography~\cite{kb,Schultz:2004kx} and muon tomography~\cite{wc}. 

The present work is motivated by our endeavor in a different field of interest,
 the measurement of reactor antineutrinos at short distances which has the 
potential to search for the possible existence of the sterile neutrino.  The 
Indian Scintillator Matrix for Reactor Anti-Neutrino (ISMRAN) is an 
experimental setup that has been developed for detecting electron antineutrinos
 emanating from the DHRUVA research reactor facility
at Bhabha Atomic Research Centre (BARC), Trombay, India using the 
Inverse beta decay (IBD) process. In the IBD process,
electron antineutrinos interact with protons of the detector resulting in
positron and neutron. The ISMRAN matrix will be used to reconstruct both
 the positron and the neutron by measuring the time-correlated prompt and
 delayed signals, respectively. The cosmogenic background, especially the
 cosmic muons generated neutrons are important background for such a 
measurement. The present work reports on the cosmic muons measurements
 using an array of plastic scintillators (PS) of the ISMRAN detector.
 We have used the PS array to measure the muon lifetime, zenith angle
 dependence of flux, and energy of muons deposited in the detector volume
 and energy spectra of its decay products.  In the next section, we briefly
 describe the ISMRAN detector setup followed by the experimental procedure
 for measurements of cosmic muons. The Monte Carlo simulation using GEANT4
 carried out to determine the detector angular acceptance and energy 
reconstruction due to cosmic muons is discussed in Sec.4. The procedure
 for data analysis is discussed in Sec.5. Finally, the results are
 summarized in the last section.
\section{The ISMRAN setup}
\begin{figure*}[]
\centering
{\includegraphics[trim={2cm 2.5cm 1.8cm 0.2cm},clip=true,width=0.6\textwidth,height=0.5\textwidth]{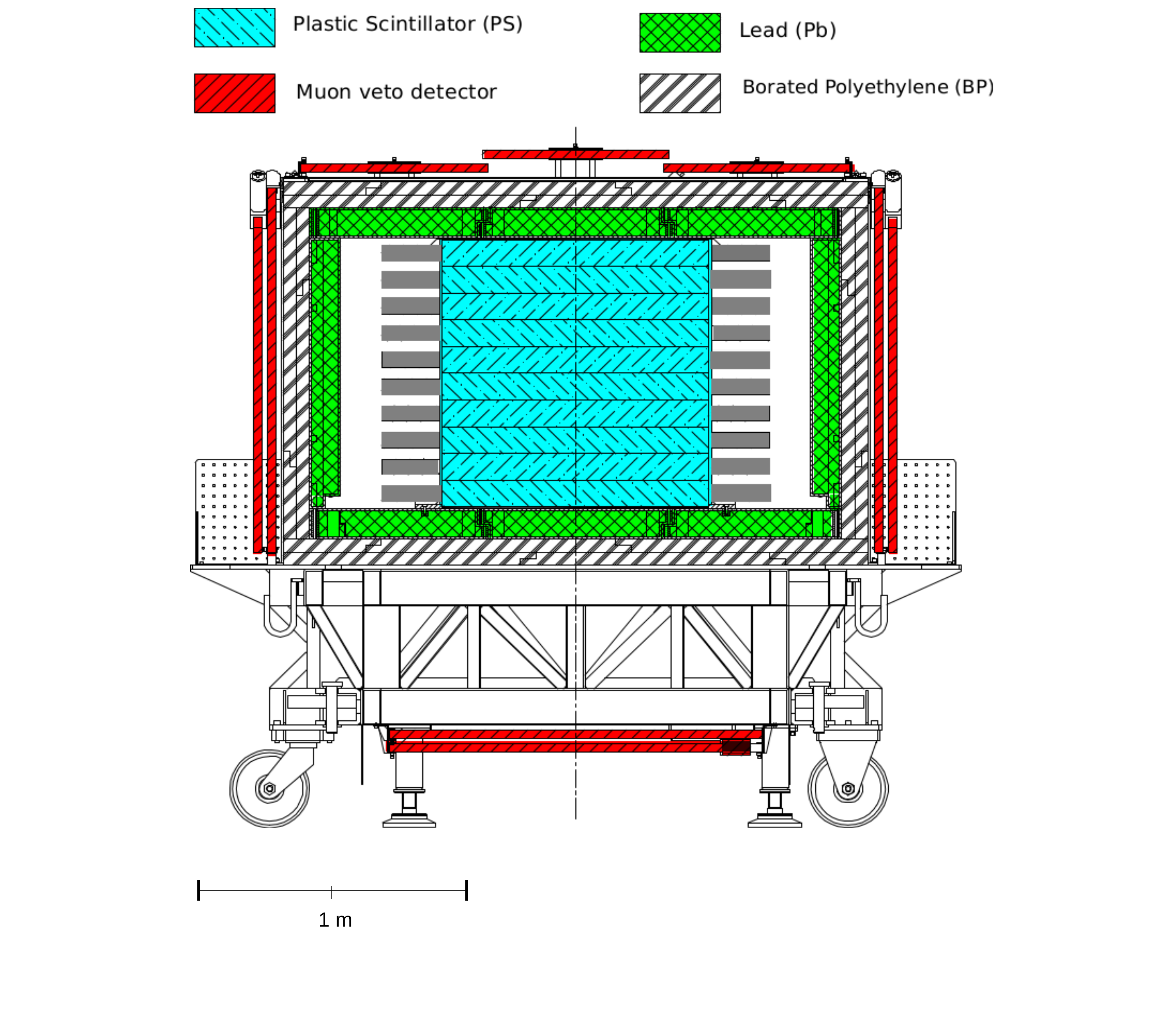}}
\hspace{-0.5cm}
{\includegraphics[trim={0.3cm 0.8cm 0cm 0cm},clip=true,width=0.42\linewidth,height=0.36\linewidth]{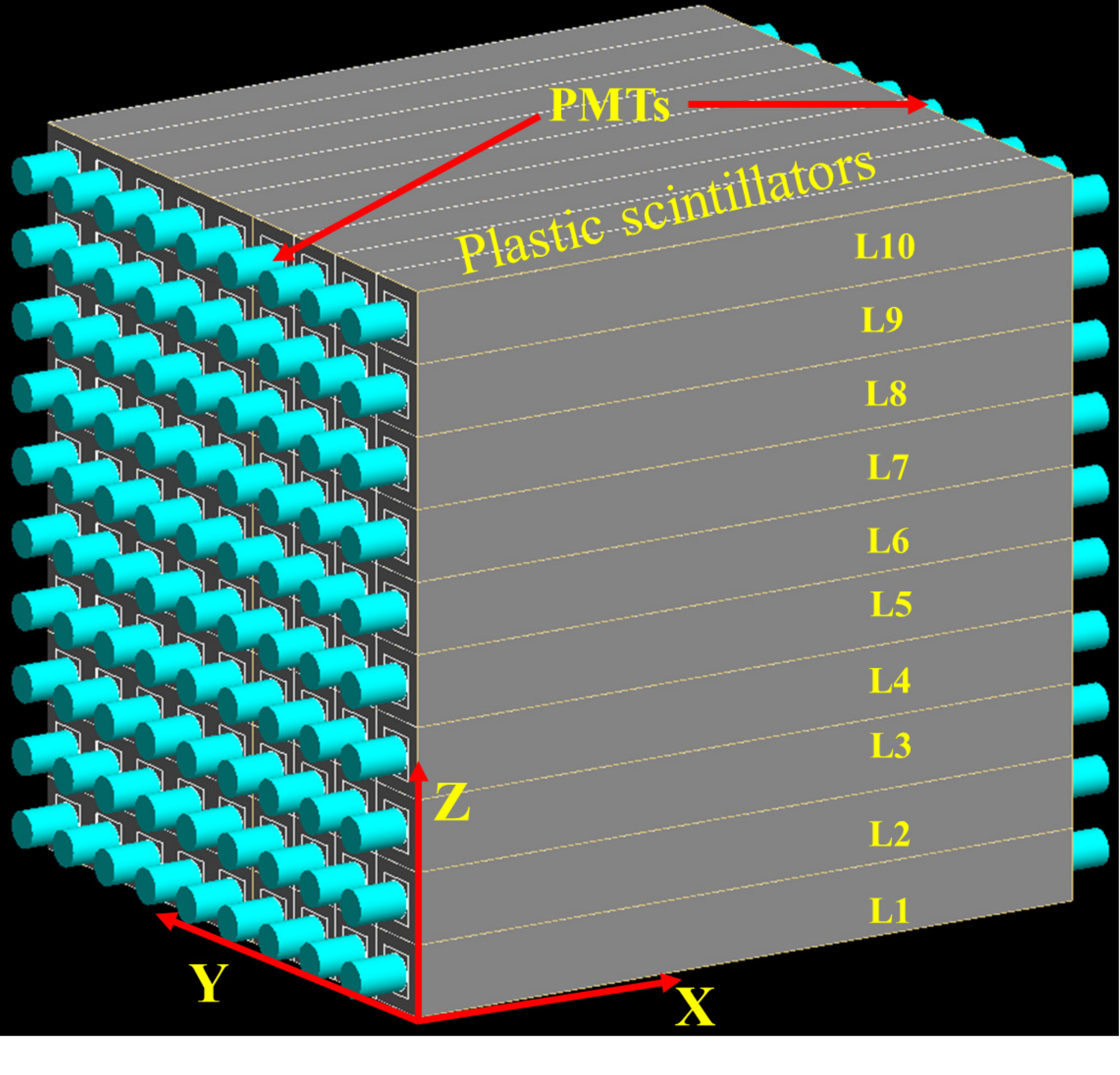}}
\caption{The left panel shows schematic of the ISMRAN setup with shielding structure and mounted on a movable base structure will be used for reactor antineutrinos measurements~\cite{Mulmule:2018efw}. The right panel shows the schematic of 9$\times$10 scintillator matrix for the measurement of cosmic muons. The
cylindrical geometry at both ends represents the PMTs coupled to the individual PS bar. }
\label{fig:schematic}
 \end{figure*}

The schematic of ISMRAN setup along with shielding structure
 and mounted on a movable base structure is shown in left panel of Fig.~\ref{fig:schematic}.
The ISMRAN setup consists of 90 PS bars each with dimensions
100 cm $\times$ 10 cm $\times$ 10 cm. 
The 9 $\times$ 10 matrix configuration of
 PS is shown in right panel of Fig.~\ref{fig:schematic}.  For the capture of
neutrons generated in the IBD process these scintillators are
 wrapped with Gadolinium ($Gd_{2}O_{3}$) coated aluminized
Mylar foils. The integrated weight of 90 such PS is close
to 1 ton. Each PS bar is coupled with two 3 inches diameter
Photo-multiplier tubes (PMT) at both ends. The PS bars are
EJ200~\cite{eljen} and PMTs used are ETL 9821 series tubes~\cite{etl}. The ISMRAN
setup will be placed above the ground at a distance of 13 m 
from the reactor core inside the reactor hall.
The detection of antineutrino events in the reactor environment is dominated by 
backgrounds of various types. There are mainly two sources of backgrounds
 namely $\gamma$-rays and neutrons from the reactor and those from natural
 backgrounds. The natural background constitutes the $\gamma$-rays mainly
 from $^{40}$K and $^{208}$Tl and spallation neutrons due to cosmic muons.
 To shield the PSs from such background, the ISMRAN active volume is surrounded 
with passive shielding material of 10 cm thick lead followed by 10 cm 
thick of borated polyethylene to reduce $\gamma$-rays and neutrons,
 respectively. Further, the set-up will be surrounded by 1-inch thick plastic
 scintillator plates from outside to veto cosmic muons or some of the cosmic
 muon-generated particles. The ISMRAN setup will be mounted on a movable 
trolley such that it can be easily maneuvered from one place to another
 to make measurements at variable distances. The schematic for the
 ISMRAN is shown in Fig.~\ref{fig:schematic}. The results presented in the current work are based
 on data acquired using the ISMRAN PS matrix without any shielding at a site away
 from the reactor in a so-called non-reactor environment.  
\section{Simulation with GEANT4} 
Monte-Carlo (MC) simulation studies are carried out using the GEANT4 (version 4.10.3) 
~\cite{Agostinelli:2002hh} simulation package which incorporated the
interaction of muons with the detector medium. Plastic scintillator detectors 
array is defined using GEANT4 detector geometry definition by including the different
 material properties through which cosmic muons have to traverse. The standard
 electromagnetic processes for $\gamma$-rays, e$^{\pm}$, and $\mu^{\pm}$ and the
QGSP BIC HP physics list for hadronic interactions are used in the simulation. In the
 present study, we use the CRY MC generator~\cite{cry} to generate cosmic muons.
For the energy and angular distribution of cosmic muons at sea level, 
CRY follows Gaisser's parametrization with modification at lower energies
 and higher zenith angles.  The generator is interfaced with a standard GEANT4. 
Muon vertices are generated randomly in 1 m $\times$ 1 m area over the detectors.
In GEANT4, it is possible to get the particle information including type, charge, position,
 momentum, and energy. 
 The hit positions on the top and bottom layers of PS bars provide the incoming 
and passing muons tracks, respectively. In the simulation, the hit positions are recorded 
from the event and particle identification information, for a particular muon path 
on an event-by-event basis. The energy loss by muons in each PS bar is also recorded. 
\section{Measurements}
The measurement of cosmic muons zenith angle, lifetime, and energy are carried out
using the PS array. The energy and position information of an event 
are extracted from the pulse height and timing information of signals
taken from PMTs of the PS bars coupled at two ends. The detector setup
 has 180 readout channels through which
PMT signals are passed to compact digitizers based data acquisition (DAQ) system.
 The modules CAEN V1730 with 16 channels and  500 MS/s frequency VME 
based waveform digitizers are used for pulse processing and acquiring the data.
\begin{figure*}[]
\centering
{\includegraphics[trim={7cm 2cm 3cm 34cm},clip=true,width=0.45\linewidth,height=0.45\linewidth]{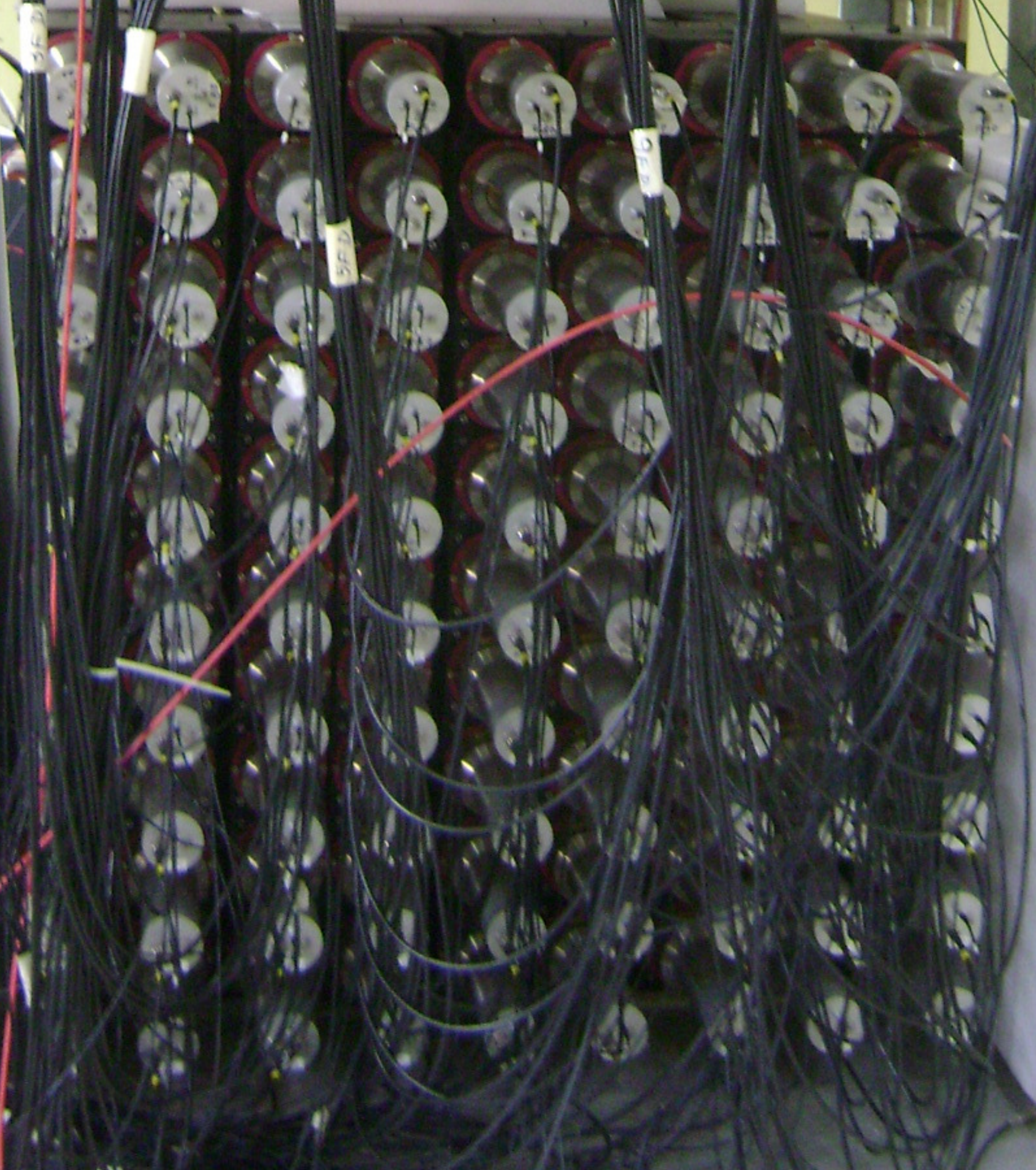}}
\hspace{0.3cm}
{\includegraphics[trim={7cm 2cm 3cm 34cm},clip=true,width=0.45\linewidth,height=0.45\linewidth]{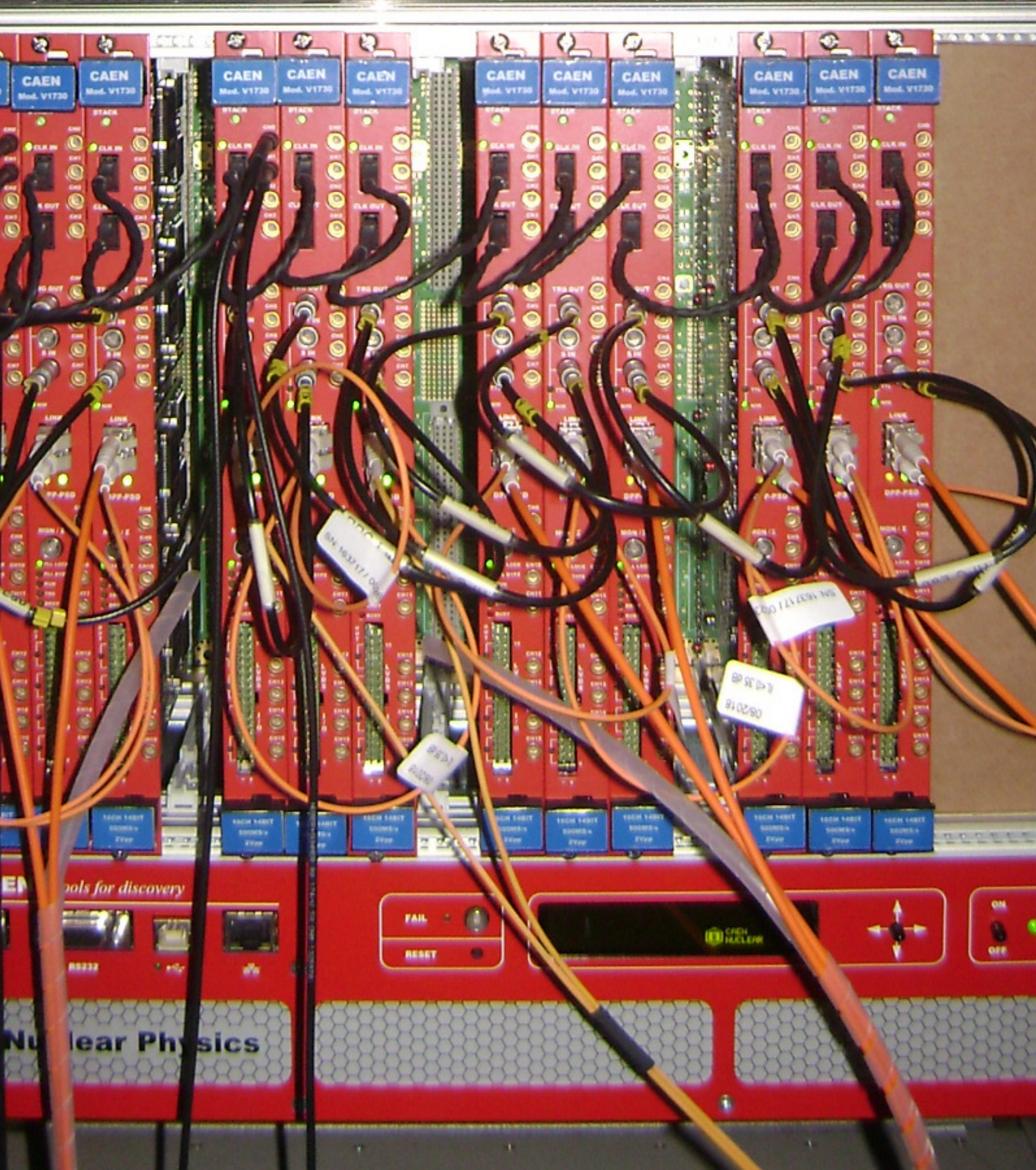}}
   \caption{\label{fig:expt}The left panel shows the PS array used for the
 present cosmic muons measurement in a non-reactor environment. The right panel shows the digitizer based data acquisition system.}
 \label{fig:matrix}
    \end{figure*}
In a single digitizer module, the discrimination, gate generation, charge integration, timing 
of each channel are processed by field-programmable gate arrays (FPGAs).  Optical communication and 
digitizing the wave-forms on board itself allows the acquisition of data to be performed at
 100 MHz rate. In comparison, the cosmic muons rate on the surface of the earth is about
 1 per cm$^2$ per min which is very small compared to the rate of data acquisition which leads
 to almost zero dead time. Events are recorded by applying a minimum operating threshold of
 200 keV which is kept the same for each channel due to the requirements of the minimum
 operating threshold of the signal as planned for the final measurements with the ISMRAN.
 The signals from PMTs are time-stamped by in-built discriminators on the digitizer board. 
In addition, the coincidence of signals from two PMTs coupled to the same PS bar is also carried 
out by FPGA. The coincidence mode results in a considerable reduction of random
background rates of acquired data. Based on the trigger efficiency study, it is
found that the optimum time window for triggering the coincidence events
from two PMT of single PS is around 16 ns~\cite{Mulmule:2018efw}. However,
in the present measurement, we have kept the coincidence time window between
two PMT signals as 36 ns as there are time offsets that may vary from one bar to another.
There are 12 digitizers used for acquiring the data for the whole matrix and all the 
digitizer modules are synchronized for time-stamps with a common reference time. 
The charge integration range of gate generated for obtaining the
 signal in ADC units of the digitizer is chosen to be 150 ns spanning the complete signal duration.
 The digitizers are programmed with the latest CAEN DPP PSD firmware version 4.11.139.6
and the working principle has been established using standard scintillator detector~\cite{rawat}. 
This firmware also allows for the discrimination of different types of radiations such as $\gamma$-rays,
 neutrons, or alpha particles if their pulse profiles in the given detector volume are different. 
The timing of each PS bar is calibrated by placing a
$\gamma$-rays $^{137}$Cs (0.662 MeV) source at different position along the
PS bar and the corresponding position-time correlation have been
obtained. Details of the time measurements have  been discussed in
Ref.~\cite{Mulmule:2018efw}. The energy calibration, as well as
the gain matching of each PS bar are carried out with known
radioactive $\gamma$-rays sources such as $^{137}$Cs (0.662 MeV),
$^{22}$Na (0.511 MeV,1.2 MeV), and AmBe (4.4 MeV).\\
Cosmic-rays measurements are carried out using the ISMRAN detector
setup as shown in the right panel of Fig.~\ref{fig:schematic} where
each PS bar is positioned in the oblong geometry. The right panel shows the
digitizer-based data acquisition system used for measurement. In the scintillator
detector, muons are identified by observing energy spectra of each
PS bar. Muons lose their energy through the ionization process while
passing through the materials. The average energy loss per unit path
length~($\frac{dE}{dx}$), for any charged particle passing through a
matter can be obtained by applying the Bethe-Bloch formula.

\begin{figure*}[]
\centering
{\includegraphics[width=0.48\linewidth,height=0.4\linewidth]{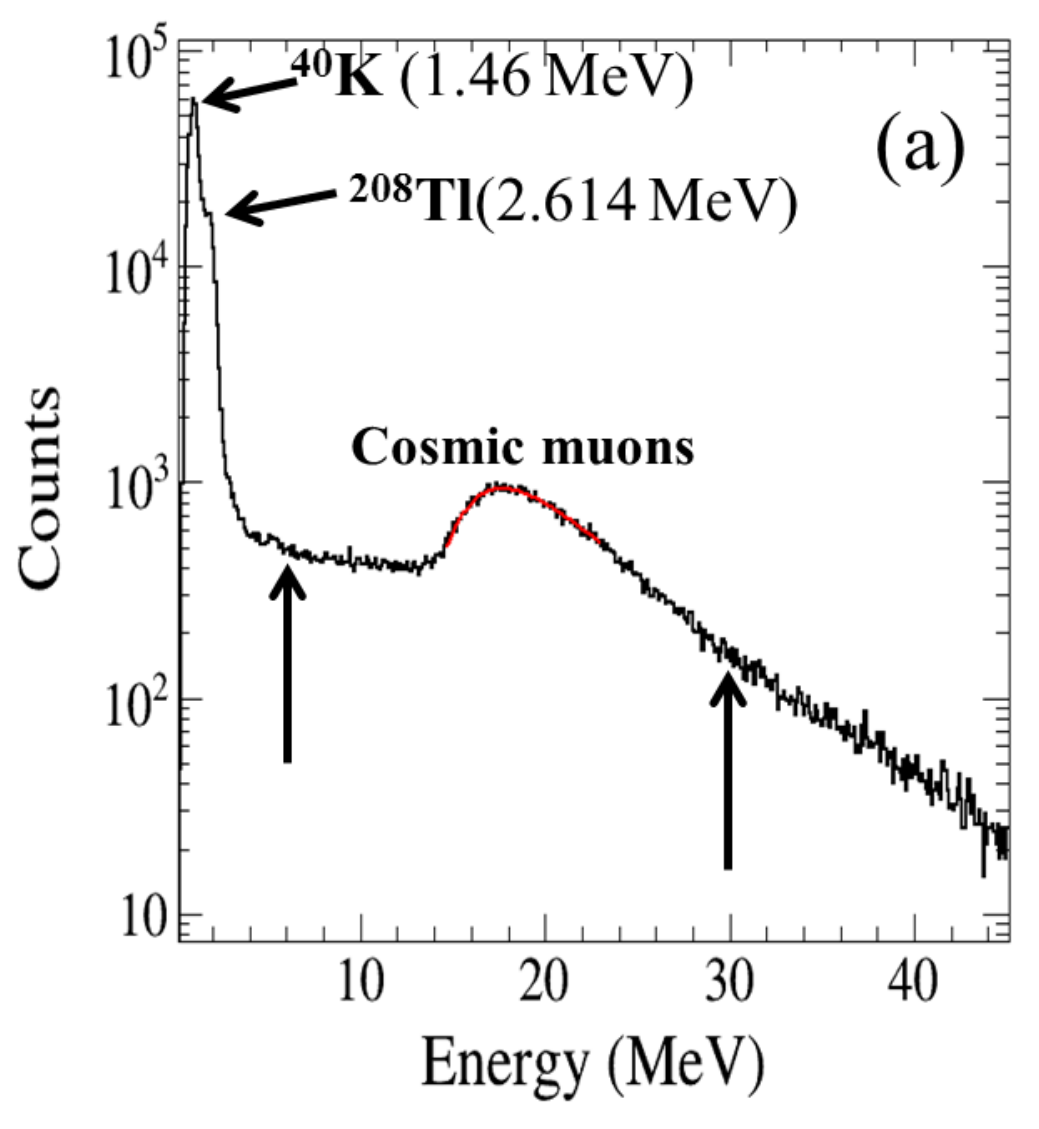}}
{\includegraphics[width=0.48\linewidth,height=0.4\linewidth]{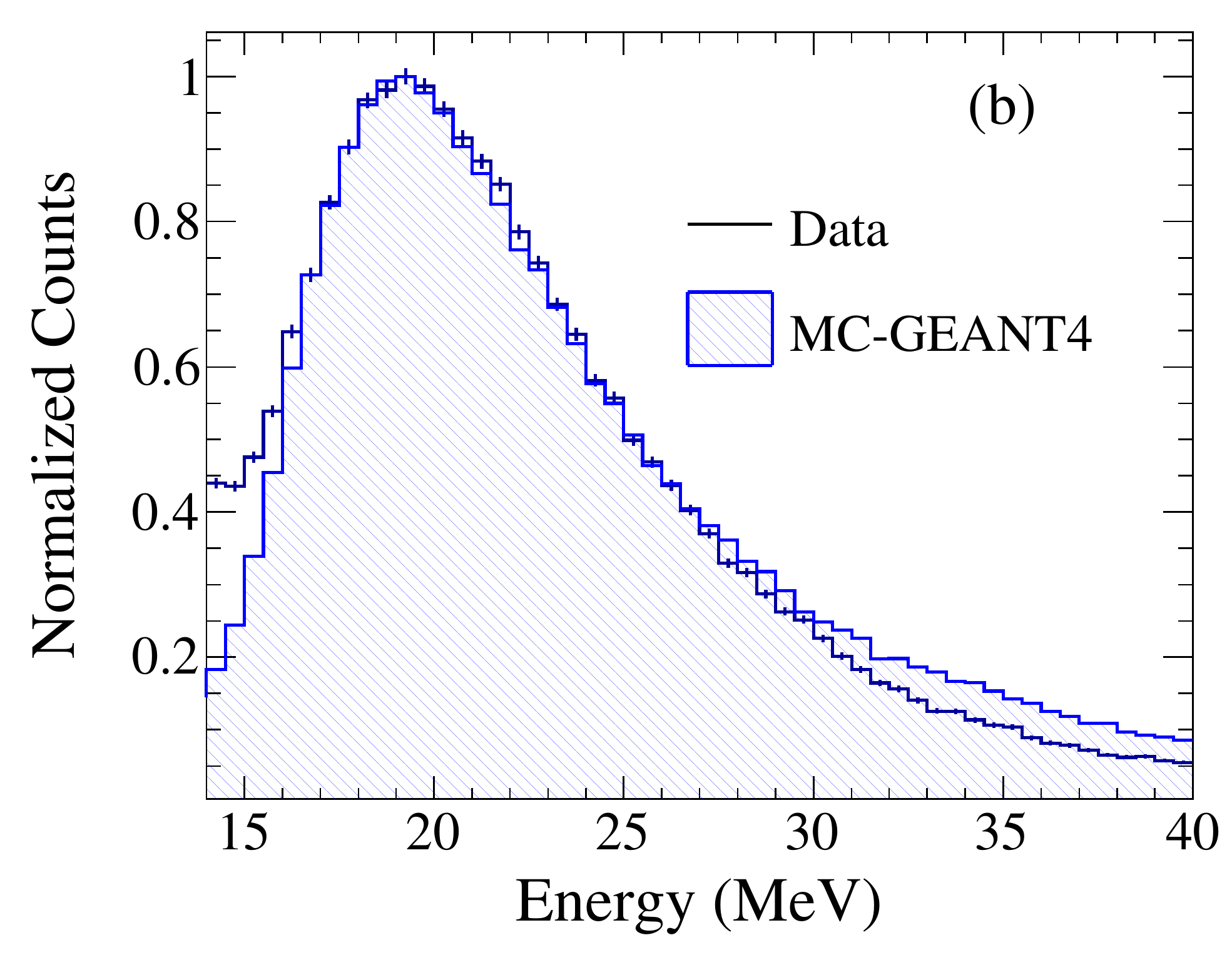}}
\caption{\label{fig:expt}(a)The measurement of energy spectrum of cosmogenic
  backgrounds with a single PS bar. (b) Simulation versus measurement of
  the energy spectra due to cosmic muons}
    \label{fig:muenergy}
    \end{figure*}

The amount of energy loss for muons passing perpendicularly to the
surface of the 10 cm thick scintillator is about 20 MeV corresponding
to the mean energy loss. The measured energy spectra of a single PS bar
are shown in the left panel of Fig.~\ref{fig:muenergy}. At low energy, structure due to
the environmental background from $^{40}$K(1.460 MeV)
 and $^{208}$Tl(2.614 MeV) is seen in the energy spectrum. The bump around 20
MeV is due to the energy deposition of cosmic muons. The right panel of
Fig.~\ref{fig:muenergy} shows the comparison between measured and
simulated energy spectra of individual PS bars due to cosmic muons.
The simulated energy spectrum is folded with detector resolution (assuming
a Landau distribution of energy spectrum) $\sim 20\%/E$. Both the measured
and simulated energy spectra match except
at lower as well as higher energies. The disagreement may be due to some of the contribution
of natural backgrounds which is absent in the simulation.
 Since energy is determined using the calibration parameters
 deduced from the low energy, it may lead to an offset, giving a difference between data and simulation. 
We have used the energy spectra, timestamp, and $X-Y$ position information
to extract various parameters like zenith-angle, lifetime, deposited energy of
stopping muons in the detector volume, and energy of charged lepton (e$^{\pm}$) due to decaying muons.
\section{Analysis of experiment data}
Angular dependence of cosmic muons reaching the detector is 
measured using the ISMRAN setup with 9$\times$10 PS matrix as shown in
right panel of Fig.~\ref{fig:schematic}. The experimental data are recorded by the DAQ system
as described above which provides the time and energy information.
The hit position of the muons are reconstructed from the time information. 
The X-positions are extracted from the time-position correlation of PS bars.
The horizontal arrangement in the matrix from left to right provides the
Y-position and the vertical layer number provide the Z-position as shown in Fig.~\ref{fig:schematic}. The position information
of the bars is used for the measurement of zenith angle-dependent muons flux.
The minimum number of hit points used for reconstructing the valid muon tracks is six.
The hit positions are fitted with a straight line in 3D. The angle of incoming muons are
extracted by taking the vector products between a fixed position (0,0,1)
related to the zenith axis and the position information of the hit. Similarly,
the sum energy of muons before decay and also the energy of the decay
products $i.e.$ e$^{+}$/e$^{-}$ are reconstructed 
using the number of bars and energy deposited in each PS bar.
 The energy threshold applied for each PS bar is the same as mentioned earlier. 
 Muons are produced randomly in the atmosphere due to the decay of
 mesons such as pions and kaons. Hence, it is expected that the time distribution of passing muons
 will follow the exponential distribution function~\cite{nima}. The time interval 
 between two consecutive muons 
reaching on the detector surface of area 1.0 $\times$ 0.9 m$^2$ 
is shown in Fig.~\ref{fig:interval}. We have selected muons
with the deposited energy in the range of 6-30 MeV as shown in Fig.~\ref{fig:muenergy}a.
The time interval distribution is fitted with an exponential function mentioned in Eq.~\ref{eq:decay}.
\begin{equation}
\label{eq:decay} 
N = N_{0}~e^{-t/\tau} + C.
\end{equation}
Here, $t$ is the time interval between the consecutive muons reaching at the detector, $\tau$ is the mean
 lifetime of arrival muons, $N_0$ and $C$ are constants. The fitted distribution is shown by red
 dashed line of Fig.~\ref{fig:interval}. It is found that the mean time interval between consecutive muons in this energy range is 
11.74 $\pm$ 0.004 (sta.) $\pm$ 0.11 (sys.) ms. The systematic uncertainty is estimated by varying the 
deposited energy range in  the PS bar.
\bef[!h]
\bc
\includegraphics[width=0.5\textwidth]{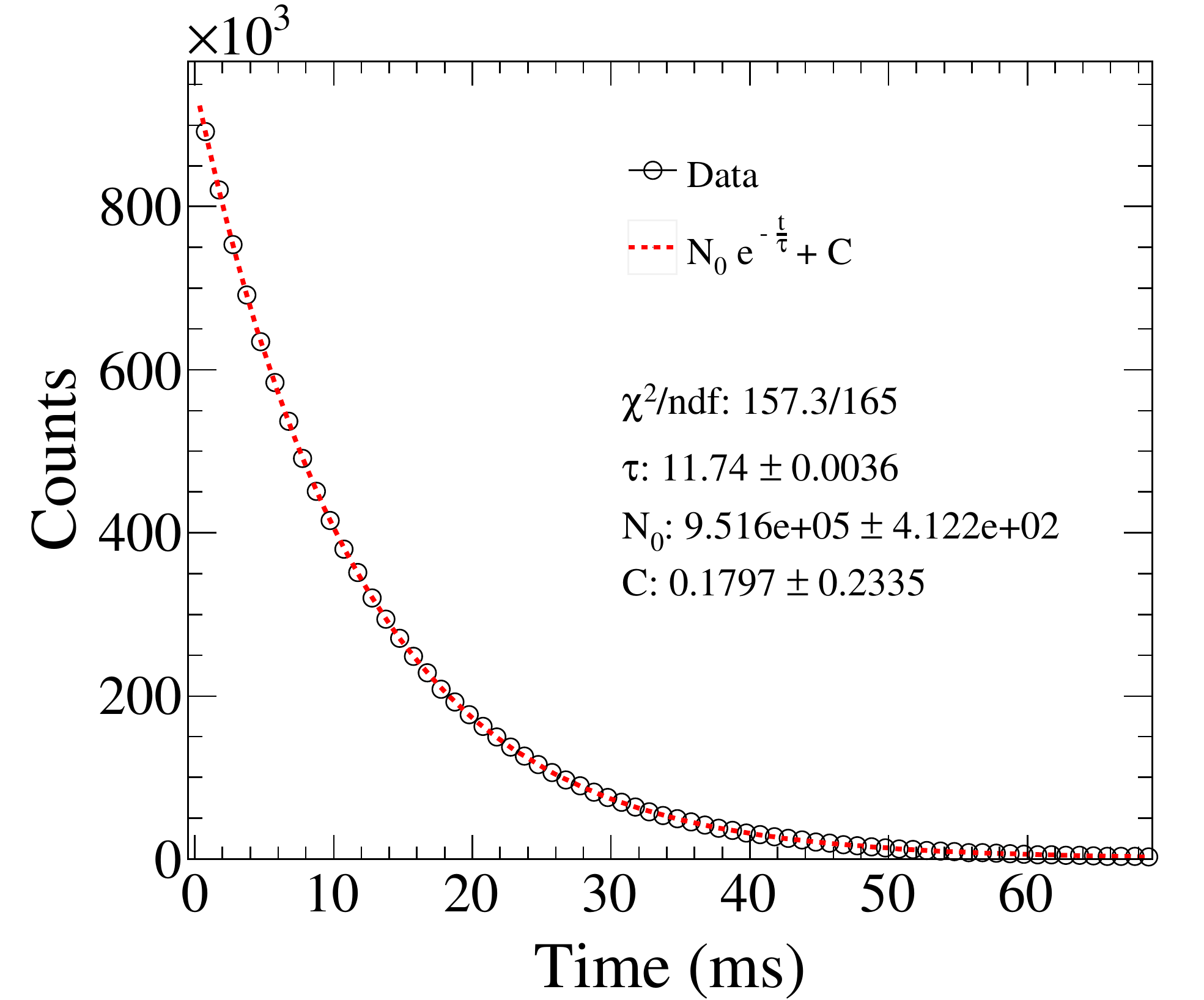}
\caption{Time interval of consecutive cosmic muons reaching on detector measured with the ISMRAN setup.}
\label{fig:interval}
\ec
\eef
\section{Results and Discussions}
The results of the measurement of zenith angle distribution,
life time, deposited energy of decaying muons, and energy of charged 
lepton as the decay product are discussed in the present section.  
\subsection{Zenith angle measurements}
Figure~\ref{fig:angle} shows the extracted zenith angle distribution of the incoming cosmic muons. 
The data points show the distribution obtained from the measurements 
which is compared with the simulation results.
 It is mentioned earlier that cosmic muons follow $cos^n \theta$ distribution.
 A good agreement between the simulation and measurement for the zenith angle
 distribution is obtained with a value of $n = 2.11$, which is comparable with the value given in Ref.~\cite{spal}.
 The variation in the predicted value of the zenith angle distribution is shown for a range of $n$ values from $1.9$ to $2.3$.
\bef[!h]
\bc
\includegraphics[width=0.5\textwidth]{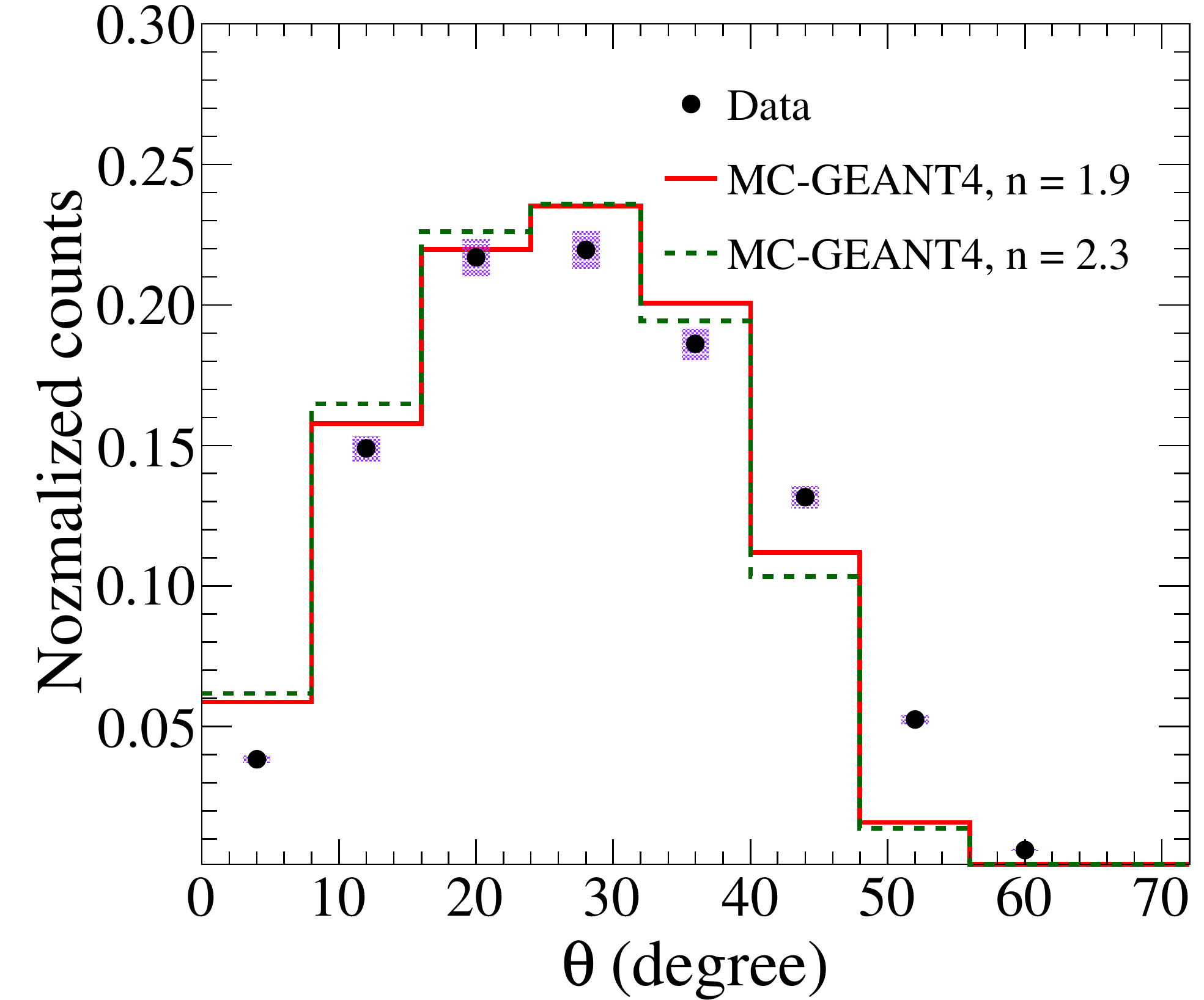}
\caption{Reconstructed zenith angle distribution of cosmic muons.}
\label{fig:angle}
\ec
\eef
The angular distribution of cosmic muons is different for various locations on the earth which depends on 
geomagnetic latitude, longitude, and altitude. There are several studies carried out by various
experimental groups~\cite{crookes, greisen,rossi,judge,fukui,gokhale,karmakar,sinha,spal,allkofer,pethu}.
 The muon flux decreases with the geomagnetic latitude while moving nearer to the equator. This is due to
 the higher values of geomagnetic cut-off rigidities near the equator. We have carried out a GEANT4 simulation  with consideration of 
concrete structure  present at the top of the experimental setup and found muons with momentum greater than 0.270 GeV/c reaching at the detector.
\begin{table}[h]
  \caption{Comparison of 
  integrated flux of cosmic muons with other experiments. P$_{\mu}$: muon momentum, P$_c$ : rigidity, S.L.: sea level, Lat.: Latitude}
  \center  
  \begin{tabular}{|l|l|l|l|l|l|}
    \hline
   
{References}& \multicolumn{2}{|c|}{Geomagnetic} & Altitude & P$_{\mu}$ 
& Integrated flux\\\cline{2-3}
&Lat.($^\circ$N)& P$_c$ (GV) & (m) & (GeV/c)& ($\times~10{^-3}~ cm^{-2}~s^{-1}~sr^{-1}$)\\

\hline
    Crookes and Rastin \cite{crookes} & 53 & 2.2  & 40 & $\geq$0.35 & 9.13 $\pm$0.12 \\
    Greisen \cite{greisen,rossi}   & 54 & 1.5 & 259 &$\geq$ 0.33 &  8.2 $\pm$0.1 \\
    Judge and Nash \cite{judge} & 53 & -- & S.L &$\geq$ 0.7 & -~- \\
    Fukui et al. \cite{fukui} & 24 & 12.6 & S.L & $\geq$ 0.34 & 7.35 $\pm$0.2 \\
    Gokhale \cite{gokhale} & 19 & -- &124 &$\geq$ 0.27  & 7.55 $\pm$0.1 \\  
    Karmakar et al \cite{karmakar} & 16 & 15.0 & 122 & $\geq$0.353 &  8.99 $\pm$0.05  \\       
    Sinha and Basu \cite{sinha} & 12 & 16.5 & 30 & $\geq$ 0.27 &  7.3 $\pm$0.2 \\    
    S.Pal \cite{spal} & 10.61 & 16 & S.L &$\geq$ 0.287 &  6.217 $\pm$0.005\\   
    Allkofer et al. \cite{allkofer} & 9 & 14.1 & S.L &  $\geq$ 0.32  & 7.25 $\pm$0.1  \\
    S. Pethuraj et al. \cite{pethu} & 1.44 & 17.6  &  160 & $\geq$ 0.11 & 7.007 $\pm$ 0.002 $\pm$ 0.526 \\
    Present data & 10.61 & 16 & S.L &$\geq$ 0.270 &  6.36 $\pm$0.002 $\pm$ 0.20\\   
    \hline
  \end{tabular}
  \label{tab:compare}
\end{table}
 We have also estimated the integral intensity of vertical incoming muons, which is given by
\begin{equation}
  \label{eqn:integral}
  I_{0} = \frac{I_{Data} }{T_{tot} \times \omega }   
\end{equation}
where, $I_{Data}$ is the total events in the observed $\theta$ distribution, $T_{tot}$ is the total time
 duration for acquiring data (in seconds), and $\omega$ is the solid angle acceptance times the 
surface area of detector setup, which is further defined as,
\begin{equation}
  \label{eqn:accp}
  \omega = \frac{A~N}{N^{\prime}}\int_{0}^{\pi/3} cos^{n}\theta~sin\theta~d\theta~\times~2~\pi
\end{equation}
where, $A$ is the surface area of the ISMRAN setup, $N$ is the number of accepted events registered in 
the bottom layer of the matrix, and $N^{\prime}$ is the number of events generated randomly on top of 
matrix using the GEANT4. It is found that the integral intensity of the vertical incoming muons
is $I_{0}$ = (6.36 $\pm$ 0.002(stat) $\pm$ 0.20 (sys.) ) $\times$ 10$^{-3}$  $cm^{-2} s^{-1} sr^{-1}$.
The systematic uncertainties are estimated by varying the deposited energy range in PS bar and
exponent ($cos^n\theta$) of zenith angle distribution, which are shown along with the data points in Fig.~\ref{fig:angle}.
  The integrated number of vertical incoming muons measured by the present setup is compared with the other measurements
which are listed in Table~\ref{tab:compare}. It is found that the vertical incoming muons measured by the present setup matches well with 
 the number given in Ref.~\cite{spal}, as the measurements are carried out at the similar geomagnetic location. 
\subsection{Life time measurement of stopped muons in the ISMRAN}
Muons with average energy $\sim$6 GeV at the production site in the atmosphere moving at speeds close to the
speed of light and are coming randomly into the detector matrix. When they enter the plastic
 scintillator, they lose energy through 
the ionization process. The high-energy muons pass through the detector volume whereas the low-energy muons
lose energy and subsequently decay via. $\mu^- \rightarrow e^- + \nu _{\mu} + \overline\nu _e$ 
and $\mu^+ \rightarrow e^+ + \overline\nu _{\mu} + \nu _e$.
The neutral particles (anti)neutrinos have weak interaction with the matter. So with
a single PS bar, it is very difficult to detect them. The charged particles e$^{+}$/e$^{-}$
have electromagnetic interaction and can be easily detected. 
\bef[!h]
\bc
\includegraphics[width=0.5\textwidth]{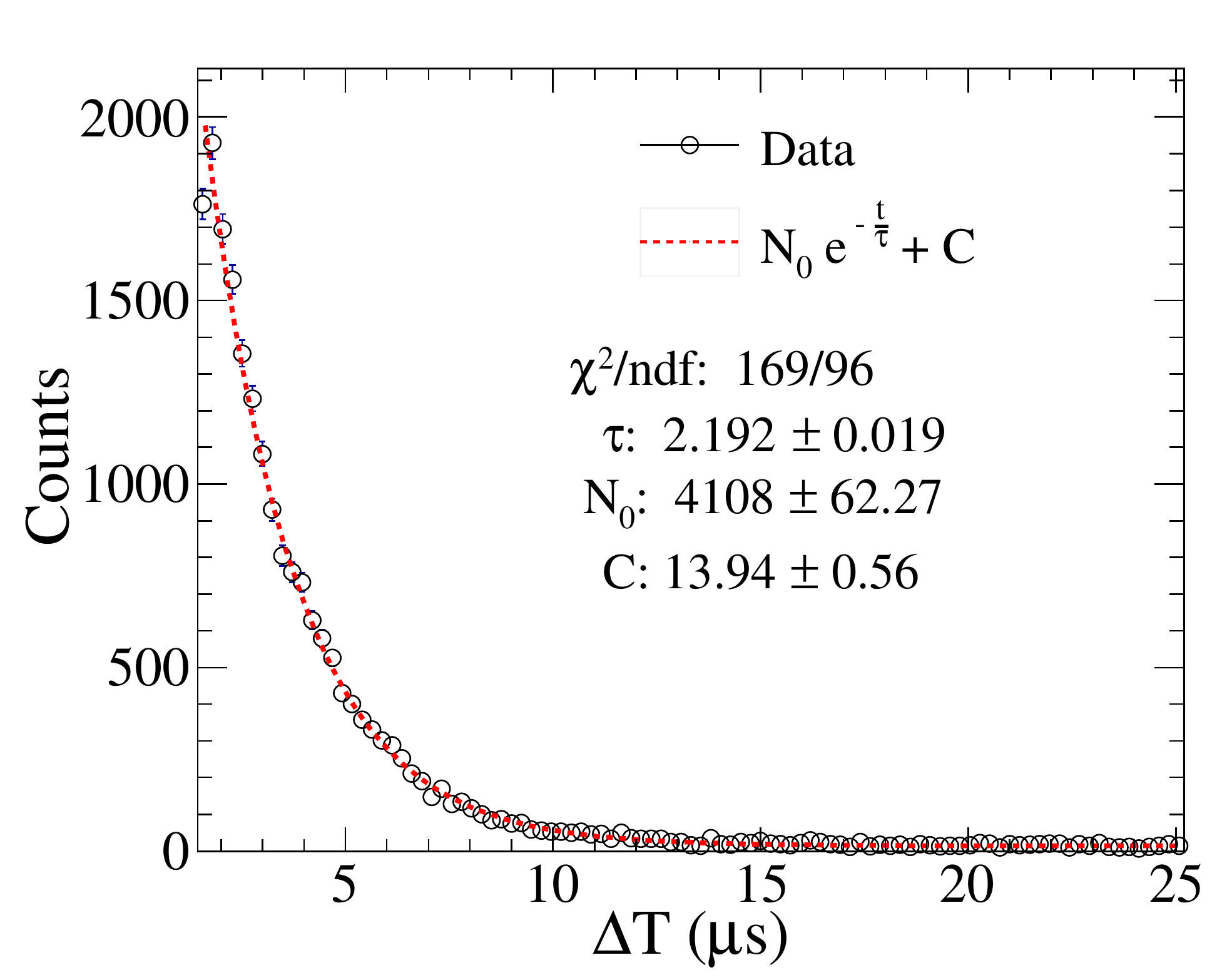}
\caption{Measured life time distribution of cosmic muons. The red line represents the distribution is fitted with a function
given in Eq.~\ref{eq:decay}.}
\label{fig:decay}
\ec
\eef
Events due to muons decay are identified by looking at two signals. The first of the signals is 
registered when muons enter the scintillator and they lose energy. This is considered as
 a prompt signal. After an interval of a few microseconds a second signal due to e$^{+}$/e$^{-}$ 
produced from $\mu^{\pm}$ decay. As mentioned earlier, signals are time-stamped, we measure the 
time difference between these correlated prompt and delayed pulses, as well as the integrated 
charge of each signal. The integrated charge is proportional to the particle energy loss in the
 scintillator detector. Muon-like events are selected by applying a cut in the deposited energy 
spectra as shown in Fig.~\ref{fig:muenergy}. The lower and upper limit for the measured energy 
spectra is 3.0 and 30.0 MeV, respectively applied on individual PS bar. The lower limit in energy
is chosen in order to reject the natural $\gamma$-rays background mainly due to $^{40}$K(1.460 MeV)
 and $^{208}$Tl(2.614 MeV). The stopped muons decaying in the detector volume can be selected 
within a selected time window which is 15 $\mu s$ in this case.
It is known that the decay of unstable nuclei and particles, including muons follow the exponential
 radioactive decay law. We have determined the time distribution
 of  decayed muons by fitting the data using the formula as given in Eq.~\ref{eq:decay}.
Here, $t$ is the time interval between the prompt and the delayed like signals due to $\mu^{\pm}$ 
and e$^{\pm}$, respectively, $\tau$ is the mean lifetime of muons, $C$ is a constant which corresponds to background from 
accidental coincidence. The time distribution of the decaying muons measured from the experiment is shown in Fig.~\ref{fig:decay} and 
it is fitted with a function given in Eq.~\ref{eq:decay}. The mean life time of muons is found as  $\tau$ = 2.192$~\pm~0.019~ \mu s$. 
\subsection{Deposited energy measurement}
The amount of energy deposited by muons or e$^{+}$/e$^{-}$  that arises due to decaying muons
in the PS bar depends upon the distance traversed by these particles in the
detector medium. More importantly, the reconstructed energy of e$^{+}$/e$^{-}$ depends
upon the decay position of muons. In order to optimize various cuts for the measured
data, we have carried out the GEANT4 simulation for studying the cosmic muon decay.
In the case of muons, the total energy loss before they decay inside the detector
volume is estimated by summing the energy deposited in the individual PS bar. A similar procedure
is also considered for reconstructing the energy of secondary charged particles such as
e$^{+}$/e$^{-}$ which are produced due to decaying muons within the detector
volume. In the process of muons decay, the maximum energy carried by electron (positron)
is 52.5 MeV~\cite{ehrlich}. The e$^\pm$ can travel up to 20 cm within the plastic scintillator
with the above maximum energy~\cite{report}. The number of PS bars fired, the total energy
deposited and, the time interval between prompt and delayed events are obtained from the
simulation are used for reconstructing total energy deposited by muons as well
as e$^{+}$/e$^{-}$ for the measured data. The energy deposited
by decayed muons and e$^{+}$/e$^{-}$ within the PS matrix are reconstructed by
applying a time window of 15 $\mu s$ between the prompt and delayed signals.
Ranges for total energy deposited are taken as 15--300 MeV for prompt and
 3--70 MeV for delayed signals. Similarly, the number of bars hit in the prompt and delayed
events are considered as 1 $< N_{hit} <$ 15 and 1 $<$ $N_{hit} <$ 6, respectively.
\begin{figure*}[]
\centering
{\includegraphics[width=0.48\linewidth,height=0.4\linewidth]{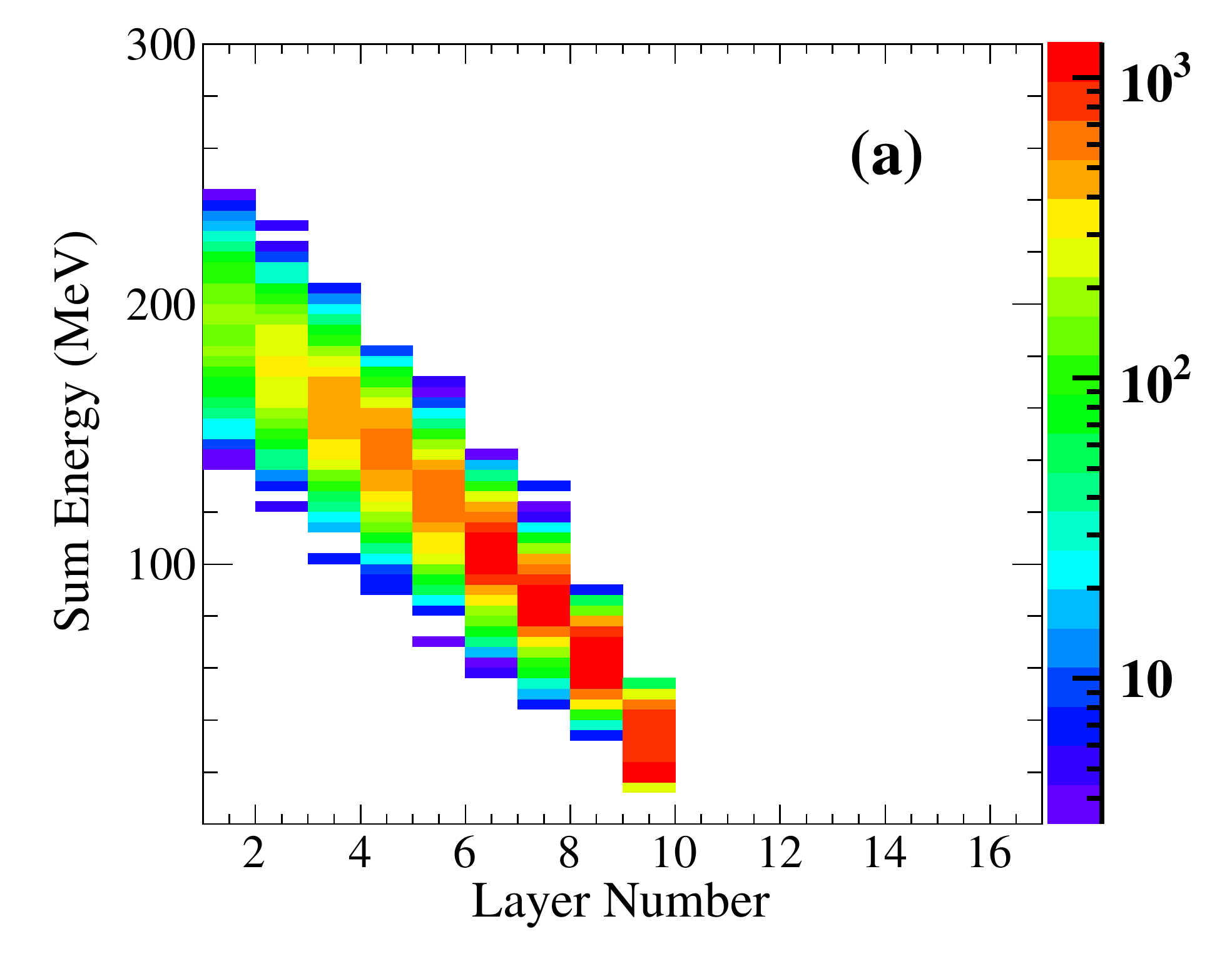}}
{\includegraphics[width=0.48\linewidth,height=0.4\linewidth]{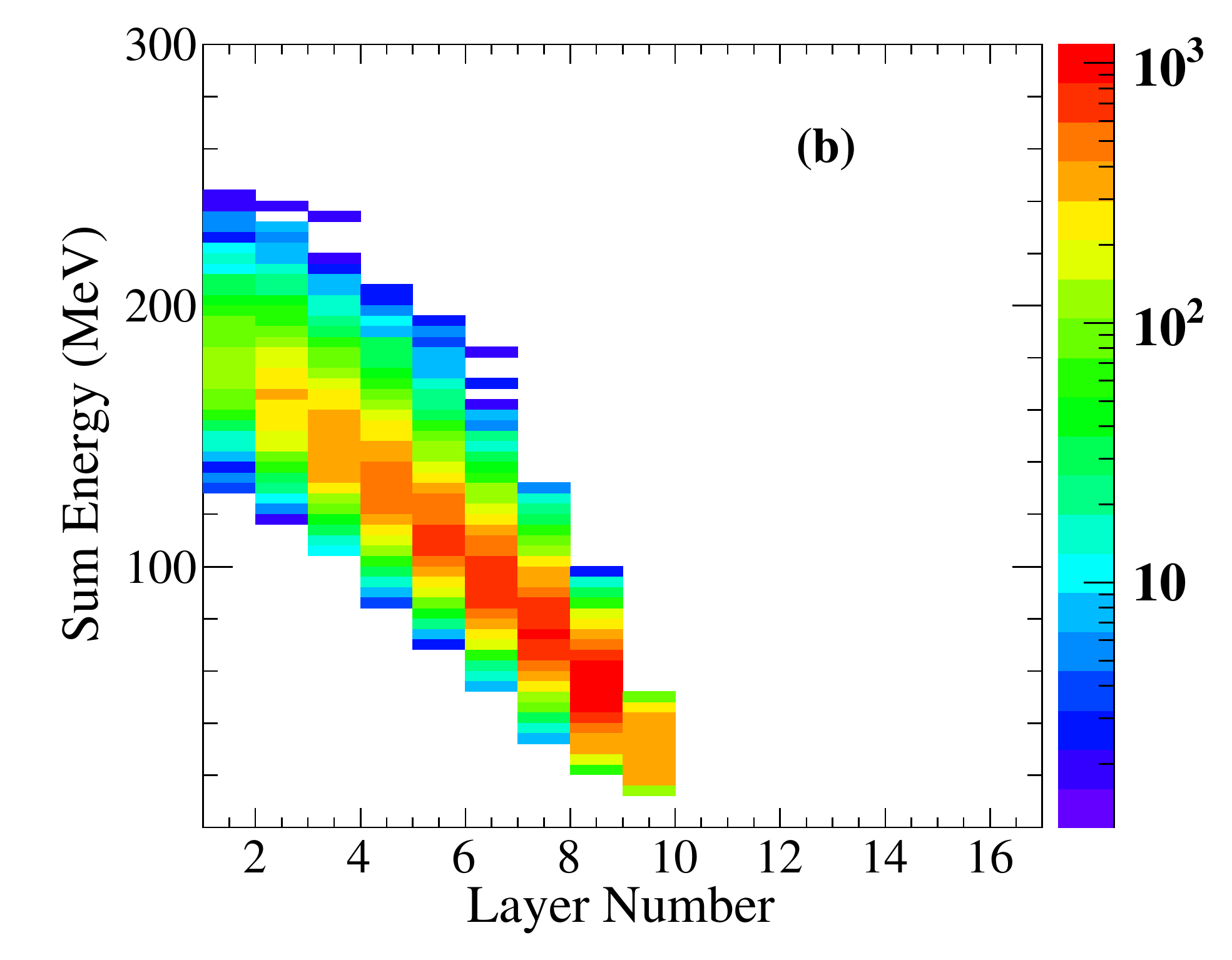}}
\caption{\label{fig:expt}Reconstruction of sum energy of decay
  muons with layer number of PS matrix (a) from GEANT4 simulation,
  (b) from the measured data.}
    \label{fig:layerdecay}
    \end{figure*}
\bef[!h]
\bc
\includegraphics[width=0.5\textwidth]{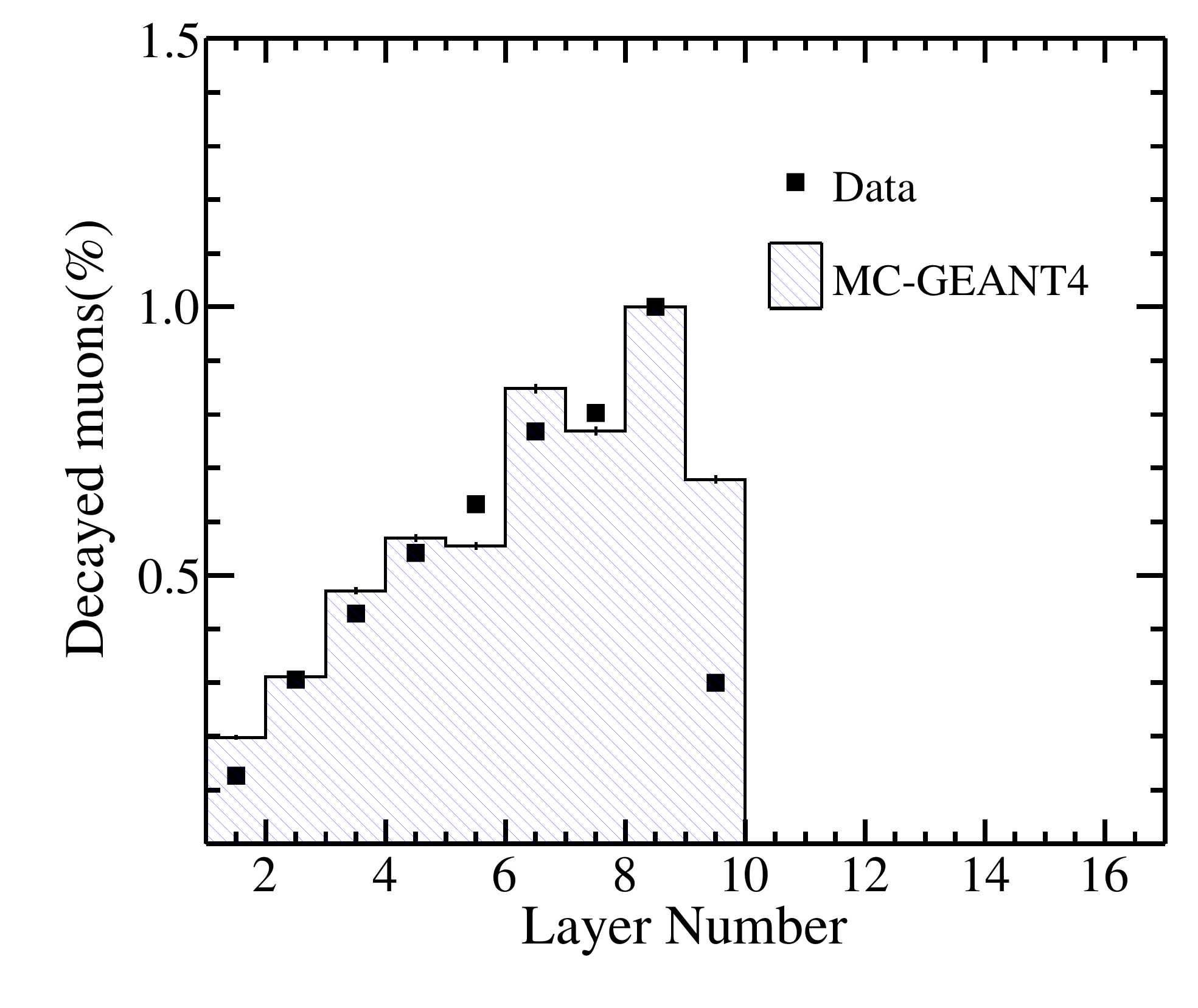}
\caption{Measured fractional distribution of cosmic muons in different layer of ISMRAN setup compared with the GEANT4 simulation.}
\label{fig:decayfrac}
\ec
\eef
Figure~\ref{fig:layerdecay}(a) shows the results obtained from simulation
for the number of muons decayed in different layers versus the total energy
deposited before it decays in the detector volume. The energy spectra of muons and e$^{+}$/e$^{-}$ are
folded with the detector resolution. In the right panel of Fig.~\ref{fig:schematic},
 the layer numbers of the detector are counted from bottom to top.
It is found from the simulation that most of the muons lose energy
within 2 to 3 layers from the top and a small fraction of muons reach at the
bottom before decaying. Similar behavior is also observed in the measured data as shown in
Fig.~\ref{fig:layerdecay}(b). Figure~\ref{fig:decayfrac} shows the comparison
between projected 1D spectra of simulated and measured data for the fraction
of muons decayed in each layer. Both simulated and measured data agree
well except for the top layer. One of the reason for disagreement may be that most 
of the low energy muons decay within the concrete structure before reaching the top
 layer of the detector setup.

The comparison of the total deposited energy by muons before decaying within
the detector volume is shown in Fig.~\ref{fig:expt}(a). The comparison of 
measured and simulated total energy spectra of e$^{+}$/e$^{-}$ are shown in 
Fig.~\ref{fig:expt}(b). Both the measured and simulated energy spectra 
match very well. In the present measurement, we have reconstructed the decaying muons from the ISMRAN setup
by measuring the time-correlated prompt and delayed signals due to muons and their decay products. 
\begin{figure*}[]
\centering
{\includegraphics[width=0.48\linewidth,height=0.4\linewidth]{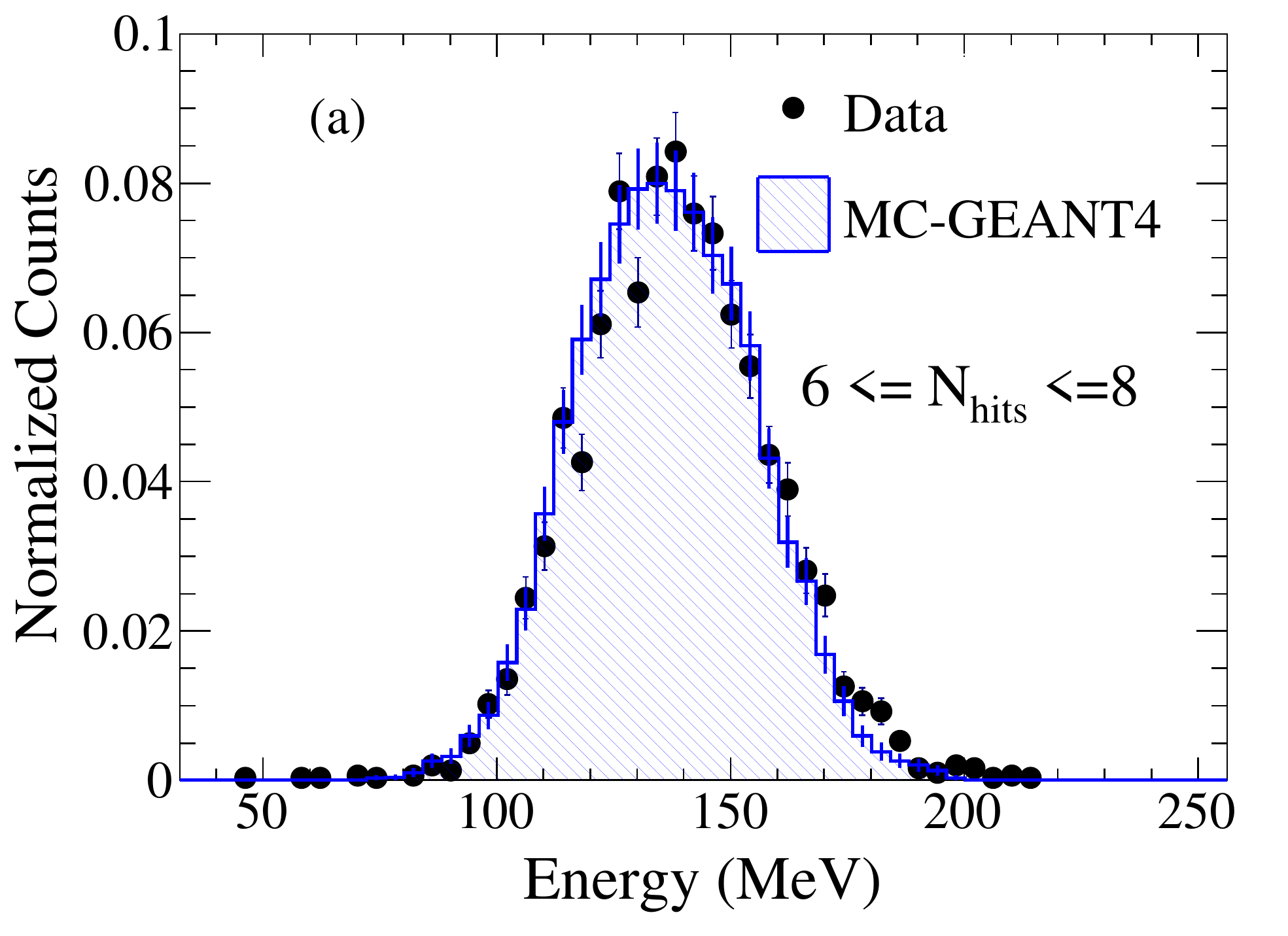}}
{\includegraphics[width=0.48\linewidth,height=0.4\linewidth]{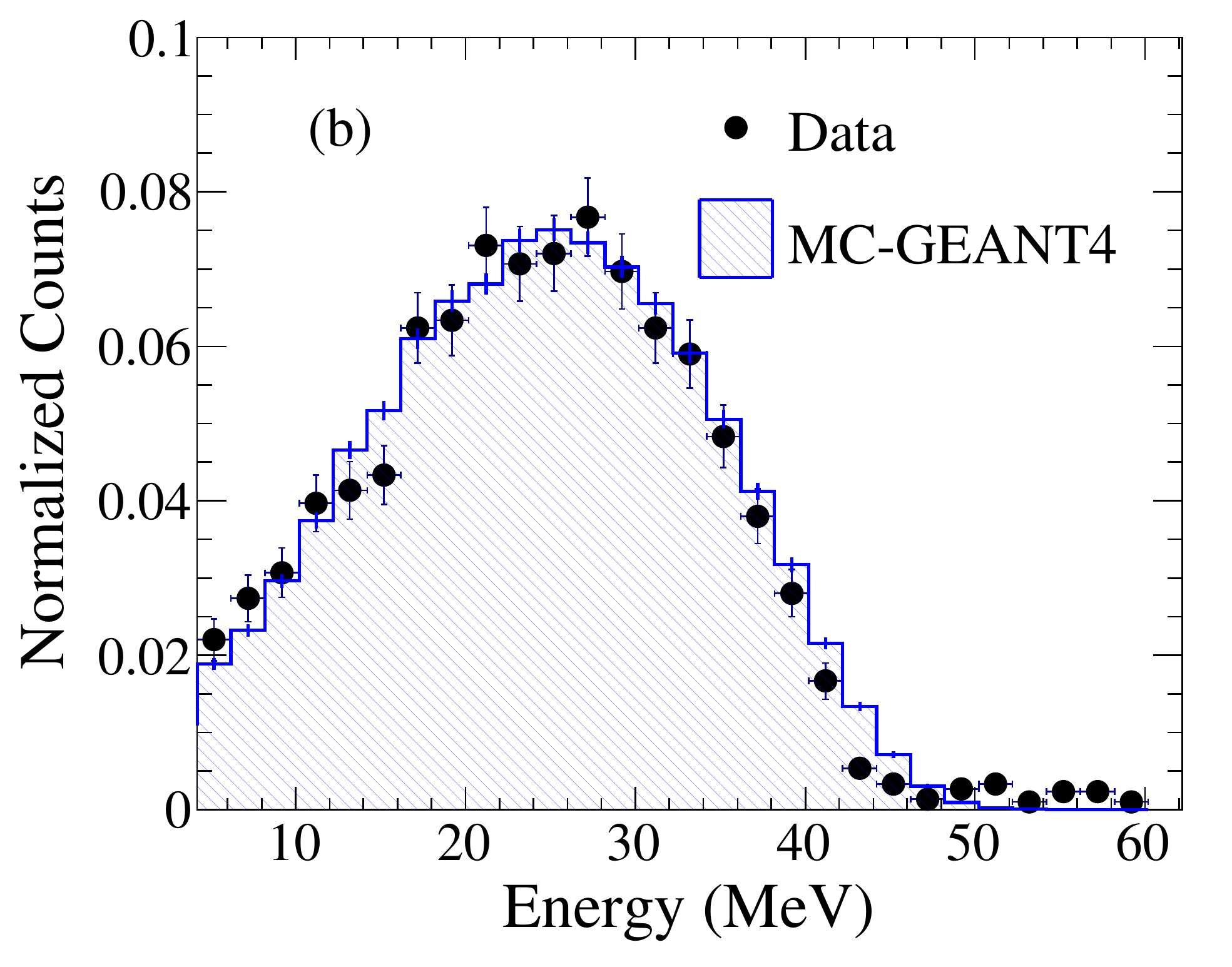}}
\caption{\label{fig:expt}(a)The energy distribution of stopped $\mu^{\pm}$
  within the detectors array (b)The energy distribution of $e^{\pm}$ due to
  decaying muons within the detectors array.}
    \label{fig:expt}
    \end{figure*}
\section{Summary}
An array of plastic scintillator detectors have been constructed to measure the 
electron antineutrinos produced from the reactors. While developing the ISMRAN setup,
 detectors are characterized and used for background measurements in a non-reactor 
environment. We have measured the cosmic muons using the ISMRAN detector setup.
 Offline data analysis has been carried out by applying suitable cuts on the number of PS bars,
 the deposited energy due to muons that decay inside the detector volume, and its decay products.
 These events which correspond to the decays of muons are selected by 
applying a time cut on prompt and delayed signals. The zenith angle 
distribution of cosmic muon flux is measured and compared with the MC results. The integral intensity of 
vertical incoming cosmic muons is measured and compared with other previous measurements.
 The detector setup was further used to measure the muon life-time and deposited energy of the 
stopped muons. The data collected serve as a useful benchmark of cosmogenic background
 in a non-reactor environment for the future measurements of electron-antineutrinos 
using the ISMRAN. After the completion of the measurements at the non-reactor environment, 
the ISMRAN setup will be installed at the Dhruva research reactor hall for the measurements
of the reactor antineutrinos.

\section*{ACKNOWLEDGMENTS}
We thank D. Mulmule for his support during the assembly of the detector setup.

\newpage

\end{document}